\documentclass[aps,preprint]{revtex4}%
\usepackage{amsfonts}
\usepackage{amsmath}
\usepackage{amssymb}
\usepackage{graphicx}%
\begin{document}
\title[Nearly integrable systems]{Level statistics for nearly integrable systems}
\author{A. Y. Abul-Magd}
\affiliation{Faculty of Engineering Sciences, Sinai University, El-Arish, Egypt}
\keywords{one two three}
\pacs{02.50.-r, 05.40.-a, 05.45.Mt, 05.45.Tp, 03.65.-w}

\begin{abstract}
We assume that the level spectra of quantum systems in the initial phase of
transition from integrability to chaos are approximated by superpositions of
independent sequences. Each individual sequence is modeled by a random matrix
ensemble. We obtain analytical expressions for the level spacing distribution
and level number variance for such a system. These expressions are
successfully applied to the analysis of the resonance spectrum in a nearly
integrable microwave billiard.

\end{abstract}
\date{\today}
\startpage{01}
\endpage{02}
\maketitle

\noindent Random matrix theory provides a framework for describing the
statistical properties of spectra for quantum systems whose classical
counterpart is chaotic \cite{mehta,guhr}. It models the Hamiltonian of the
system by an ensemble of random matrices, subject to some general symmetry
constraints. Time-reversal-invariant quantum system are represented by a
Gaussian orthogonal ensemble (GOE) of random matrices when the system has
rotational symmetry and by a Gaussian symplectic ensemble otherwise. \ Chaotic
systems without time reversal symmetry are represented by the Gaussian unitary
ensemble (GUE). A complete discussion of the level correlations even for these
three canonical ensembles is a difficult task. \ Most of the interesting
results are obtained for the limit of large matrices. Analytical results have
long ago been obtained for the case of two-dimensional matrices \cite{porter}.
It yields simple analytical expressions for the nearest-neighbor-spacing
distribution (NNSD), renormalized to make the mean spacing equal one. The
spacing distribution for the GOE, $p(s)=\frac{\pi}{2}s\exp\left(  -\frac{\pi
}{4}s^{2}\right)  $, where $s$ is the spacing between adjacent energy levels
rescaled to unit mean spacing $D$, is known as Wigner's surmise. Analogous
expression $p(s)=\frac{32}{\pi^{2}}s^{2}\exp\left(  -\frac{4}{\pi}%
s^{2}\right)  ,$ is obtained for the GUE \cite{porter,haake}.

There are elaborate theoretical arguments by Berry and Tabor \cite{tabor} that
classically integrable systems should have Poissonian statistics. The Poisson
distribution of the regular spectra has been proved in some cases (see,
results by Sinai \cite{sinai} and Marklof \cite{marklof}, for instance). Still
its mechanism is not completely understood. It has also been confirmed by many
numerical studies, although the deviations of the calculated $P(s)$ from
$\exp(-s)$ are often statistically significant (see \cite{rob} and references
therein).\ The appearance of the Poisson distribution is now admitted as a
universal phenomenon in generic integrable quantum systems.

A typical Hamiltonian system shows a phase space in which regions of regular
motion and chaotic dynamics coexist. These systems are known as mixed systems.
Their dynamical behavior is by no means universal, as is the case for fully
regular and fully chaotic systems. If we perturb an integrable system, most of
the periodic orbits on tori with rational frequencies disappear. However, some
of these orbits persist. Elliptic periodic orbits appear surrounded by
islands. They correspond to librational motions around these periodic orbits
and reflect their stability. The Kolmogorov-Arnold-Moser (KAM) theorem states
that invariant tori with a sufficiently incommensurate frequency vector are
stable with respect to small perturbations. Numerical simulations show that
when the perturbation increases more and more tori are destroyed. For large
enough perturbations, there are locally no tori in the considered region of
phase space. The break-up of invariant tori leads to a loss of stability of
the system, that is, to chaos. There are three main scenaria of transition to
global chaos in finite-dimensional (nonextended) dynamical systems, one via a
cascade of period-doubling bifurcations, a Lorenz system-like transition via
Hopf and Shil'nikov bifurcations, and the transition to chaos via
intermittence \cite{eckmann,elnashaie,bunimovich}. It is natural to expect
that there are other (presumably many more) such scenaria in extended
(infinite-dimensional) dynamical systems.

The nature of the stochastic transition is more obscure in quantum than in
classical mechanics. So far in the literature, there is no rigorous
statistical description for the transition from integrability to chaos. The
assumptions that lead to the RMT description do not apply to mixed systems.
While some elements of the Hamiltonian of a typical mixed system could be
described as randomly distributed, the others would be non-random. Moreover,
the matrix elements need not all have the same distributions and may or may
not be correlated. Thus, the RMT approach is a difficult route to follow.
Comprehensive semiclassical computations have been carried out for Hamiltonian
quantum systems, which on the classical level have a mixed phase space
dynamics (see, e.g. \cite{gut} and references therein). Berry and Robnik
elaborated a NNSD for mixed systems based on the assumption that
semiclassically the eigenfunctions and associated Wigner distributions are
localized either in classically regular or chaotic regions in phase space
\cite{berryrobnik}. Accordingly, the sequences of eigenvalues connected with
these regions are assumed to be statistically independent, and their mean
spacing is determined by the invariant measure of the corresponding regions in
phase space. There have been several proposals for phenomenological random
matrix theories that interpolate between the Wigner-Dyson RMT and banded
random matrices with an almost Poissonian spectral statistics \cite{rosen}.
Unfortunately, these works do not lead to valid analytical results, which
makes them difficult to use in the analysis of experimental data. There are
other phenomenological approaches (see, e.g. \cite{abul} and references
therein), which use nonextensive statistical mechanics, based on maximizing
Tsallis or Kaniadakis entropies \cite{Ts,kania}, as well as the recently
proposed concept of superstatistics. These approaches have the advantage of
conserving base invariance of the Hamiltonian matrix. They provide a
satisfactory description near the end of transition from integrability to chaos.

This paper considers another phenomenological approach, which has the spirit
of the KAM theorem, to the stochastic transition in quantum systems. The phase
space of the integrable system consists of infinitely many tori corresponding
to the conserved symmetries of the system. In the semiclassical limit, energy
eigenstates are expected to be localized on individual tori. Tori destruction
corresponds to the mixing of the corresponding quantum eigenstates. Symmetry
breaking breaks some of the invariant tori but only deforms others according
to the KAM theorem. Quantum symmetry breaking \ strongly mixes a limited
number of eigenstates, but has a less influence on the other ones. Thus, the
spectrum is divided into independent sequences of eigenvalues. States
belonging to the same sequence are strongly mixed. The sequence may be
modelled by a GOE if time reversal invariance is preserved. The interaction
between states belonging to different sequences grows as symmetry breaking
increases. This amounts to amalgamating the initial sequences into a fewer
number of independent sequences with no more regular character. Consequently,
as the number of the no-symmetry sequences decreases, their fractional density
increases accordingly. As the state of chaos is reached, the whole spectrum
consists of a single (GOE) sequence.

Abul-Magd and Simbel \cite{abul1} consider another class of mixed systems, in
which the degrees of freedom are divided into two noninteracting groups, one
having chaotic dynamics and one regular. The Hamiltonian of such a system is
given as a sum of two terms, so that each of the eigenvalues of the total
Hamiltonian is expressed as a superposition of two eigenvalues corresponding
to the two Hamiltonian terms. The spectrum is then given by a superposition of
independent chaotic subspectra. Each subspectum corresponds to one (or one
set) of the quantum numbers of the regular component of the Hamiltonian. This
model is used in \cite{abul2} to describe level statistics of vibrational
nuclei. An elaborated version of this model \cite{dembo} has been applied to
study NNSD of a wide range of nuclei \cite{ahsw}.

We shall now consider the energy spectra of nearly integrable systems that may
be represented as a superposition of independent sequences $S_{i}$ each having
fractional level density $f_{i}$ , with $i=1,...,m$, and with $\sum
\limits_{i=1}^{m}$ $f_{i}$ $=1$. In this case, NNSD of\ the composite spectrum
can be exactly expressed in terms NNSD's of the constituting sequences (see,
e.g., Appendix A.2 of Mehta's book \cite{mehta}). The gap probability
function
\begin{equation}
E(s)=\int_{s}^{\infty}ds^{\prime}\int_{s^{\prime}}^{\infty}ds"p(s")
\end{equation}
that gives the probability of finding no eigenvalues in segment of length
$s~$of the total spectrum, is expressed as a product of the gap functions of
the individual sequences%
\begin{equation}
E(m,s)=%
{\displaystyle\prod\limits_{i=1}^{m}}
E_{i}(f_{i}s).
\end{equation}
We assume that all of $S_{i}$ obey the GOE statistics. Then NNSD of each of
the individual sequences distribution is given by the Wigner surmise, then for
all $i$%
\begin{equation}
E_{i}(x)=E_{\text{GOE}}(x)=\text{Erfc}\left(  \frac{\sqrt{\pi}}{2}x\right)  ,
\end{equation}
where Erfc$\left(  x\right)  $\ is the complimentary error function. The NNSD
of the full spectrum can the be obtained by twice differentiating the
resulting gap function.

We shall restrict our consideration to the case when all sequences have the
same fractional level density, so that%
\begin{equation}
f_{i}=f=1/m.
\end{equation}
The gap function of the composite spectrum is then given by%
\begin{equation}
E(m,s)=\left[  \text{Erfc}\left(  \frac{\sqrt{\pi}}{2m}s\right)  \right]
^{m}.
\end{equation}
Differentiating this function twice with respect to $s$, we obtain%
\begin{equation}
p(m,s)=\left[  \text{Erfc}\left(  \frac{\sqrt{\pi}}{2m}s\right)  \right]
^{m-2}e^{-\pi s^{2}/4m^{2}}\left[  \left(  1-\frac{1}{m}\right)  e^{-\pi
s^{2}/4m^{2}}+\frac{\pi s}{2m^{2}}\text{Erfc}\left(  \frac{\sqrt{\pi}}%
{2m}s\right)  \right]  . \label{mGOE}%
\end{equation}
It is easy to see that, for a single sequence $p(1,s)=\frac{\pi}{2}%
s\exp\left(  -\frac{\pi}{4}s^{2}\right)  $, so that the Wigner surmise is
recovered. On the other hand, $\lim_{m\rightarrow\infty}p(m,s)=e^{-s}$\ as required.-

A weak point of the distribution in Eq. \ref{mGOE} is that it differs from
zero at $s$ = 0, because the symmetry-breaking interaction lifts the
degeneracies. The model thus fails in the domain of small spacings as far as
the NNSD's are concerned. The magnitude of this domain depends on the ratio of
the strength of the symmetry-breaking interaction to the mean level spacing.
Therefore, it is expected to work well for nearly integrable system.

In the case when the individual sequences are described by a GUE, the gap
function for each individual sequence is given by%
\begin{equation}
E_{\text{GUE}}(x)=e^{-4x^{2}/\pi}-x\text{ Erfc}\left(  \frac{2}{\sqrt{\pi}%
}x\right)  .
\end{equation}
In this case, the NNSD of the composite spectrum is given by%
\begin{multline}
p(m,s)=\frac{1}{m^{3}}\left[  e^{-4s^{2}/\pi m^{2}}-\frac{s}{m}\text{Erfc}%
\left(  \frac{2s}{m\sqrt{\pi}}\right)  \right]  ^{m-2}\label{mGUE}\\
\times\left\{  \left(  m-1\right)  \left[  \frac{4s}{\pi}e^{-4s^{2}/\pi m^{2}%
}+m~\text{Erfc}\left(  \frac{2s}{m\sqrt{\pi}}\right)  \right]  ^{2}\right. \\
+\left.  \frac{32s^{2}}{\pi^{2}}e^{-4s^{2}/\pi m^{2}}\left[  e^{-4s^{2}/\pi
m^{2}}-\frac{s}{m}\text{Erfc}\left(  \frac{2s}{m\sqrt{\pi}}\right)  \right]
\right\}  .
\end{multline}

The situation with the level number variance (LNV) $\Sigma^{2}$ of composite
spectra is not as clear as in the case of NNSD. Seligman and Verbaarschot
\cite{seligman} argued that $\Sigma^{2}$ is a variance and can therefore be
expressed for a composite spectrum as a sum of the corresponding quantities
for its subspectra,%
\begin{equation}
\Sigma^{2}(m,L)=\sum_{i=1}^{m}\Sigma_{i}^{2}(f_{i}L),
\end{equation}
where $\Sigma_{i}^{2}(x)$ is the LNV of the $i$th sequence. There, the LNV of
the composite spectrum composed of $m$ independent sequences described by RMT
is given by%
\begin{equation}
\Sigma^{2}(m,L)=m~\Sigma_{\text{RMT}}^{2}(L/m) \label{Sig}%
\end{equation}
where $\Sigma_{\text{RMT}}^{2}(L)$ is the LNV calculate by RMT. Explicit
expressions for $\Sigma_{\text{RMT}}^{2}(L)$ in the cases of GOE and GUE are
given in Mehta's book.

We shall compare our predictions for the NNSD and the LNV with the energy
spectra of a Lima\c{c}on billiard. This is a closed billiard whose boundary is
defined by the quadratic conformal map of the unit circle $z$ to $w$,
\begin{equation}
w=z+\lambda z^{2},|z|=1.
\end{equation}
The shape of the billiard is controlled by a single parameter $\lambda$. For
$0\leq\lambda<1/4$, the Lima\c{c}on billiard has a continuous and convex
boundary with a strictly positive curvature and a collection of caustics near
the boundary. At $\lambda=1/4$, the boundary has zero curvature at its point
of intersection with the negative real axis, which turns into a discontinuity
for $\lambda>1/4$. The classical dynamics of this system and the corresponding
quantum billiard have been extensively investigated by Robnik and
collaborators \cite{robnik}. They concluded that the dynamics in the
Lima\c{c}on billiard undergoes a smooth transition from integrable motion at
$\lambda=0$ via a soft chaos KAM regime for $0<\lambda\leq1/4$ to a strongly
chaotic dynamics for $\lambda=1/2$. We assume that the quantum dynamics of the
Lima\c{c}on billiard can approximately be described by the model present here.
The spherical billiard for which $\lambda$ = 0 has two good quantum numbers,
namely the energy and angular momentum. As $\lambda$ increases, the spherical
symmetry is gradually destroyed. States corresponding to different
angular-momentum quantum numbers mix to different degrees depending on the
magnitude of the wavefunctions at large $z$.

The resonance spectra in microwave cavities with the shape of billiards from
the family of Lima\c{c}on billiards have been constructed for the values
$\lambda={0.125,0.150,0.300}$ and the first 1163, 1173 and 942 eigenvalues
were measured, respectively \cite{rehfeld, hesse}. The billiard with
$\lambda=0.300$ has a chaotic dynamics and its resonance spectrum is well
described by a GOE \cite{DA}, i.e. using Eqs. \ref{mGOE} and \ref{Sig} with
$m=1$. We here consider the $\lambda={0.125,0.150}$ billiards that exhibit
mixed regular-chaotic dynamics, which is predominantly regular. We have
performed a least-square analysis of the NNSD and LNV for these billiards
using Eqs. \ref{mGOE} and \ref{Sig}, respectively, taking $m$ as a real
parameter. The best fit values are $m=3.21,2.62$ for the billiards with
$\lambda=0.125,0.150$, respectively. Figure 1 shows the result of comparison
of the experimental NNSD for these billiards with Eq. \ref{mGOE}, while the
result for the LNV is given in FIG 2. We note,however, the interpretation of
$m$ as the number of spectra that are being superimposed suggests that it
should be an integer. For this reason, we show in the two figures the results
of calculation with 3 and 4 sequences for the $\lambda=0.125$ billiard and 2
and 3 sequences for the $\lambda=0.150$ one. The figures show that the
agreement with the fractional value of $m$ is not so much better than with the
integer values of the parameter. In both cases, the parameter $m$ can be taken
equal to 3 for both the NNSD and LNV in spite of the fact that the NNSD is
close to a Poisson distribution while the LNV shows a large amount of spectral
rigidity. This unusual situation is in favor of the validity of the present
model. To demonstrate this we show an analysis of the NNSD in FIG 3 and of the
LNV in FIG 4 using the Berry-Robnik model \cite{berryrobnik}. The best-fit
value of the parameter $q$ that measures the fractional volume of the regular
part of the phase space is found for the NNSD to be 0.585. This quite
different from the value 0.156 that fits the LNV. We note that the agreement
between the prediction of the Berry-Robnik model and the experimental LNV is
worse than the agreement with our model. Concerning the NNSD, both models fail
to describe the depletion in the number of events in the first bin. There is
100 spacings in this bin, so that the statistical error is 10 \%. Thus the
depletion is statistically significant. The disagreement reflects the partial
neglect of level repulsion in both model where the superimposed sequences are
considered as independent.

The expression for NNSD of a spectrum composed of independent sequences with
non-equal fractional densities $f_{i}$ is more complicated. It has been shown
in \cite{abul1,dembo} that the NNSD in this case essentially depends on a
single parameter, namely $\left\langle f\right\rangle =\sum\limits_{i=1}%
^{m}f_{i}^{2},$ which is the mean fractional level density for the
superimposed sequences; the statistical weight of each sequence is given again
by its fractional density. For a superposition of equal sequences,
$f_{i}=\left\langle f\right\rangle =1/m.$ Therefore, Eq. \ref{mGOE} can
approximately be used to describe a superposition of independent but not equal
sequences by considering $m$\ as a parameter, not necessarily taking integer
values. The non-integer parameter $m$ will play the role of an effective
number of the constituting sequences $m=1/\left\langle f\right\rangle .$ One
can adopt this interpretation of the parameter $m$ if one sees that the fit in
FIGs 1 and 2 are deteriorated by taking $m=3$ instead of 3.21 or 2.62,

To summarize, we consider a model for systems with regular-chaotic dynamics in
which the energy spectrum is represented by an independent sequences of
levels, each one modeled by a Gaussian random ensemble. By varying the
effective number of sequences, the model interpolates between the Poissonian
spectrum for the regular system where the spectrum consists of infinite number
of sequences and that of a chaotic system whose spacing distribution is
approximated by the Wigner surmise. We show that the model successfully
describe both the NNSD and LNV for a nearly integrable Lima\c{c}on billiard
with the same value of the model parameter.

\pagebreak

Figure Caption

\bigskip

FIG 1. Comparison between the NNSD for two nearly integrable Lima\c{c}on
billiards with the same distribution for $m$ independent GOE sequences.

FIG 2. Comparison between the LNV for two nearly integrable Lima\c{c}on
billiard with the same distribution for $m$ independent GOE sequences.

FIG 3. Comparison between the NNSD for a nearly integrable Lima\c{c}on
billiard ($\lambda=0.125$) with the same distribution calculated using the
Berry-Robnik semiclassical method.

FIG 4. Comparison between the LNV for a nearly integrable Lima\c{c}on billiard
($\lambda=0.125$) with the same distribution calculated using the Berry-Robnik
semiclassical method.

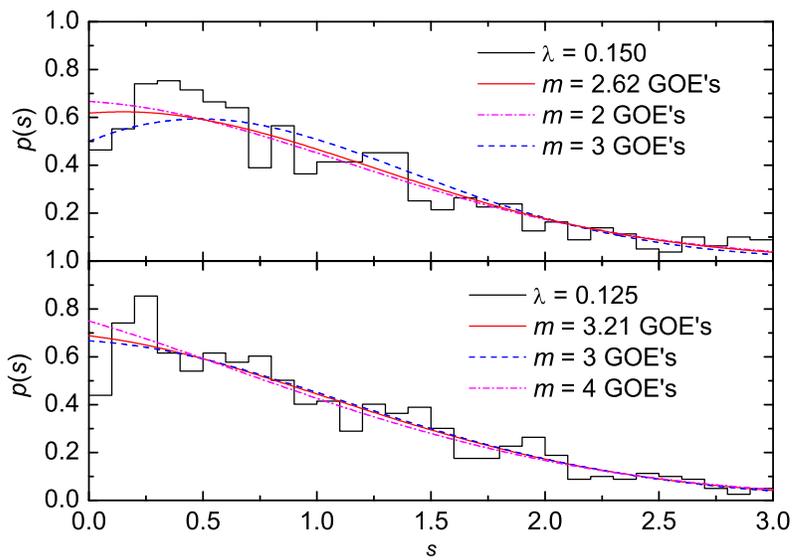
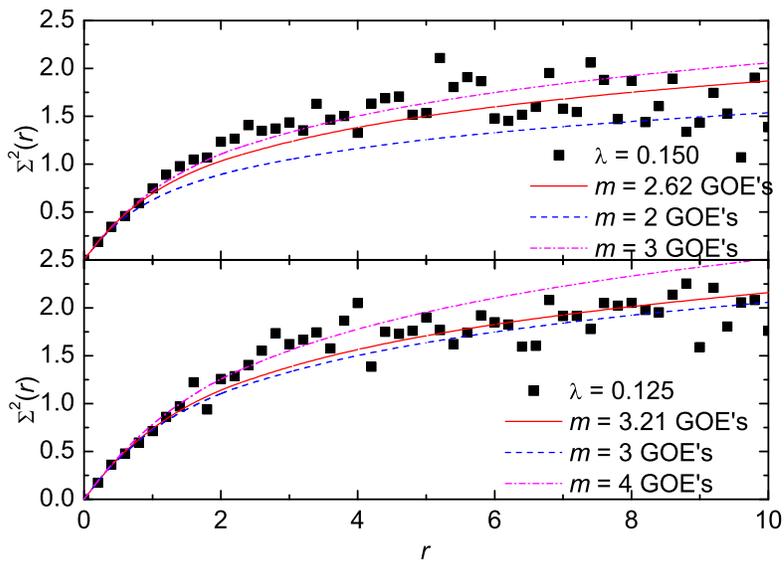
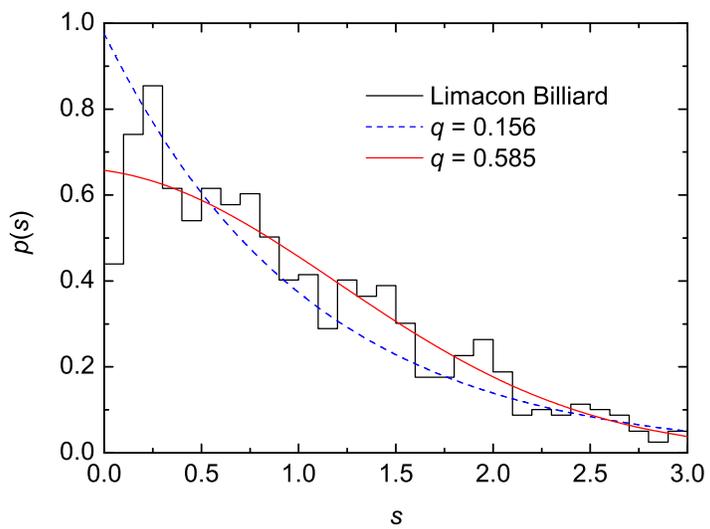
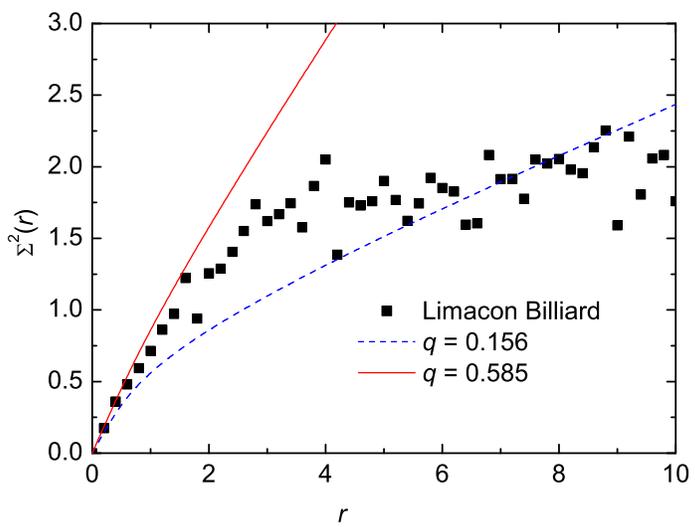
\end{document}